\documentclass[12pt,preprint]{aastex}
\title{X-ray variability during the quiescent state of the neutron-star X-ray transient in the
globular cluster NGC~6440}
\shortauthors{Cackett et al.}
\shorttitle{The Quiescent X-ray Transient in NGC 6440} 

\author{Edward M. Cackett\altaffilmark{1}, Rudy Wijnands\altaffilmark{2},
 Craig O. Heinke\altaffilmark{3}, Peter D. Edmonds\altaffilmark{3},
Walter H. G. Lewin\altaffilmark{4}, David Pooley\altaffilmark{4},
Jonathon E. Grindlay\altaffilmark{3}, Peter G. Jonker\altaffilmark{3},
Jon M. Miller\altaffilmark{3}}

\altaffiltext{1}{School of Physics \& Astronomy, University of St Andrews, North Haugh,
St Andrews, Fife, KY16 9SS, Scotland, UK; emc14@st-and.ac.uk}
\altaffiltext{2}{Astronomical Institute `Anton Pannekoek', University of Amsterdam,
Kruislaan 403, 1098 SJ, Amsterdam, The Netherlands}
\altaffiltext{3}{Harvard-Smithsonian Center for Astrophysics, 60 Garden Street,
Cambridge, MA 02138, USA}
\altaffiltext{4}{Center for Space Research, Massachusetts Institute of Technology, 77
Massachusetts Avenue, Cambridge, MA 02139, USA}

\begin{document}

\begin{abstract}
The globular cluster NGC 6440 is known to harbor a bright neutron-star X-ray
transient.  We observed the globular cluster with \textit{Chandra} on two
occasions when the bright transient was in its quiescent state in July 2000 and
June 2003 (both observations were made nearly 2 years after the end of their
preceding outbursts). The quiescent spectrum during the first observation is well
represented by a two component model (a neutron-star atmosphere model plus a
power-law component which dominates at energies above 2 keV). During the second
observation (which was roughly of equal duration to the first observation) we found
that the power-law component could no longer be detected.
Our spectral fits indicate that the effective temperature of the
neutron-star surface was consistent between the two observations.
We conclude that the effect of the change in power-law component caused the
0.5-10 keV flux to be a factor of $\sim$2 lower during the second observation
compared to the first observation.  We discuss plausible explanations for the
variations, including variable residual accretion onto the neutron star
magnetosphere or some variation in the interaction of the
pulsar wind with the matter still outflowing from the companion star.
\end{abstract}


\section{INTRODUCTION}

X-ray transients are a sub-group of low-mass X-ray binaries - binary
systems in which a compact object (either a neutron star or a black hole) is accreting
matter from a low-mass companion star.  They are normally
in a quiescent state, during which very little or no accretion onto the
compact object occurs.  Occasionally, however, these systems go into outburst, lasting weeks
to months, throughout which they increase greatly in brightness.
Such outbursts arise due to a large increase in the mass-accretion rate onto the compact object.

About a dozen neutron star X-ray transient systems have been detected in quiescence as well as
in outburst.  The quiescent luminosity of these
neutron star systems is typically $10^{32} - 10^{34}$ erg s$^{-1}$ compared to typical outburst
luminosities in the range $10^{36} - 10^{38}$ erg s$^{-1}$.  The X-ray spectra of these
quiescent neutron star systems are generally characterized by two components: a soft
component (which dominates the spectra below a few keV), equivalent to a blackbody temperature
of $kT \sim 0.1 - 0.3$ keV, and a hard
power-law X-ray tail (which dominates the spectra above a few keV).
To explain the X-ray emission of neutron star X-ray transients in their quiescent
state several models have been proposed \citep[for example, residual accretion onto the neutron 
stars, e.g.,][]{1999MNRAS.305...79M,2001ApJ...557..304M, 2000ApJ...541..849C}, but the most
widely used model is that in which the emission is due to the cooling of the neutron star
which has been heated during the outbursts \citep[see e.g.,][]{1998ApJ...504L..95B}.  In the
\citet{1998ApJ...504L..95B} model, the neutron star core is heated by
nuclear reactions occurring deep in the crust due to accretion during outburst, and this heat is
released as thermal emission during quiescence.  The X-ray emission from the cooling of a
neutron star cannot directly account for the hard power-law tail seen in the X-ray spectra and
the nature of this hard component is uncertain.  Possible explanations for the power-law tail
are accretion onto the neutron star magnetosphere or pulsar shock emission
\citep[e.g., see the discussion in][]{1998A&ARv...8..279C}.

Globular clusters are ideal for studying X-ray transients in quiescence as the distance
to the host clusters can usually be determined more accurately than for the Galactic quiescent
X-ray binaries.  Here we report on \textit{Chandra} observations of the globular cluster NGC~6440
which is known to harbor a bright neutron-star X-ray transient.
NGC~6440 is a globular cluster at a distance of $8.5 \pm 0.4$ kpc and reddened by
\textit{E(B-V)} = 1.0 \citep{1994A&AS..108..653O}.
A transient in NGC~6440 was first seen during an outburst in December 1971 to January 1972
\citep{1975Natur.257...32M,1976ApJ...207L..25F}.  In 1998 August, a second outburst was
detected by the Wide Field Camera (WFC) on-board \textit{BeppoSAX} \citep{1998IAUC.6997....3I} and
the All-Sky Monitor (ASM) on-board the \textit{Rossi X-ray Timing Explorer (RXTE)}.  During this
outburst four type I X-ray bursts were detected \citep{1999A&A...345..100I}, demonstrating the
neutron star nature of the compact object.  An optical follow-up of this outburst found two
possible candidates, V1 and V2, for the optical counterpart of the transient
\citep{2000A&A...359..960V}.

\citet{2001ApJ...563L..41I} and \citet{2002ApJ...573..184P} reported on a \textit{Chandra}
observation of NGC~6440 nearly 2 years
after the end of the 1998 outburst.  The neutron-star X-ray transient was in quiescence and
\citet{2002ApJ...573..184P}
detected 24 low-luminosity X-ray sources in this globular cluster.  They also reported that 4-5
quiescent neutron stars might be present in this cluster, based on their X-ray
luminosities and soft, thermal X-ray spectra.  Their source CX1 was the brightest low-luminosity
source in the cluster and its position was consistent with that of V1 in
\citet{2000A&A...359..960V} and possibly the quiescent counterpart of the bright transient
source.  This issue was resolved in August 2001, when the \textit{RXTE}/ASM and the
\textit{BeppoSAX}/WFC detected another outburst from the transient in NGC~6440. 
\citet{2001ApJ...563L..41I} obtained a brief Chandra observation during this outburst which
resulted in a sub-arcsecond position of the source.  They found that the 1998 and 2001 transient
is associated with CX1 from \citet{2002ApJ...573..184P} and V1 from
\citet{2000A&A...359..960V}.  From here on, we refer to the quiescent counterpart of the
neutron-star transient in NGC~6440 as CX1 after \citet{2002ApJ...573..184P}.

To study the quiescent counterpart CX1 of the neutron-star transient and the additional
low-luminosity X-ray sources in NGC~6440 in more detail, we observed NGC~6440 for a third time
using \textit{Chandra}.  Our observation was taken whilst this transient was in a quiescent state
and we compare this new \textit{Chandra} observation with the previous \textit{Chandra} 
observation of the transient in a quiescent state by \citet{2001ApJ...563L..41I} and
\citet{2002ApJ...573..184P} to study potential
variability of the quiescent X-ray emission of the source. The other globular cluster low
luminosity sources within NGC~6440 will be discussed in a future paper.

\section{OBSERVATIONS AND ANALYSIS}

On 2003 June 26 we observed NGC~6440 with \textit{Chandra} for 24 ks using only the S3 chip of
the ACIS-S detector.  In Figure \ref{fig:asm} we show the \textit{RXTE} ASM light
curve of the transient since January 1996.  In this figure we indicate when the \textit{Chandra}
observations were performed.  It can be seen that CX1 was not in a bright outburst phase during
the 2000 July 4 observation and our 2003 June 26 observation (as we shall show below, CX1 was in
a quiescent state during both observations).  The location of the August 1998 outburst is also
marked on.  This outburst is barely detected with the \textit{RXTE} ASM, and in this
figure the data are binned over 7 days and so only one bin is slightly above the noise
\citep[see also][]{1999A&A...345..100I}.  Since February 1999, the Proportional Counter Array
(PCA) on-board \textit{RXTE} has been monitoring the Galactic centre region, which includes
NGC~6440, about twice a week \citep{2001ncxa.conf...94S}.  Since the start of this monitoring
campaign, only the 2001 outburst has been detected (C. B. Markwardt 2004, private communication).
The PCA is an order of magnitude more sensitive than the ASM and so weak outbursts which might
have been missed by the ASM should have been detected by the PCA.  However, we cannot exclude
that before February 1999, weak outbursts with X-ray fluxes below the ASM detection threshold,
may have gone undetected.

Data reduction and analysis of the 2003 June 26 and 2000 July 4 observations was
performed using the CIAO 3.0.2 software package
provided by the \textit{Chandra} X-ray Center and following the threads listed on the CIAO web 
pages\footnote{Available at http://cxc.harvard.edu/ciao/}. 
Background flares were searched for but none were found so we used all available data in our
analysis.

\subsection{Image Analysis}

\begin{figure}
\plotone{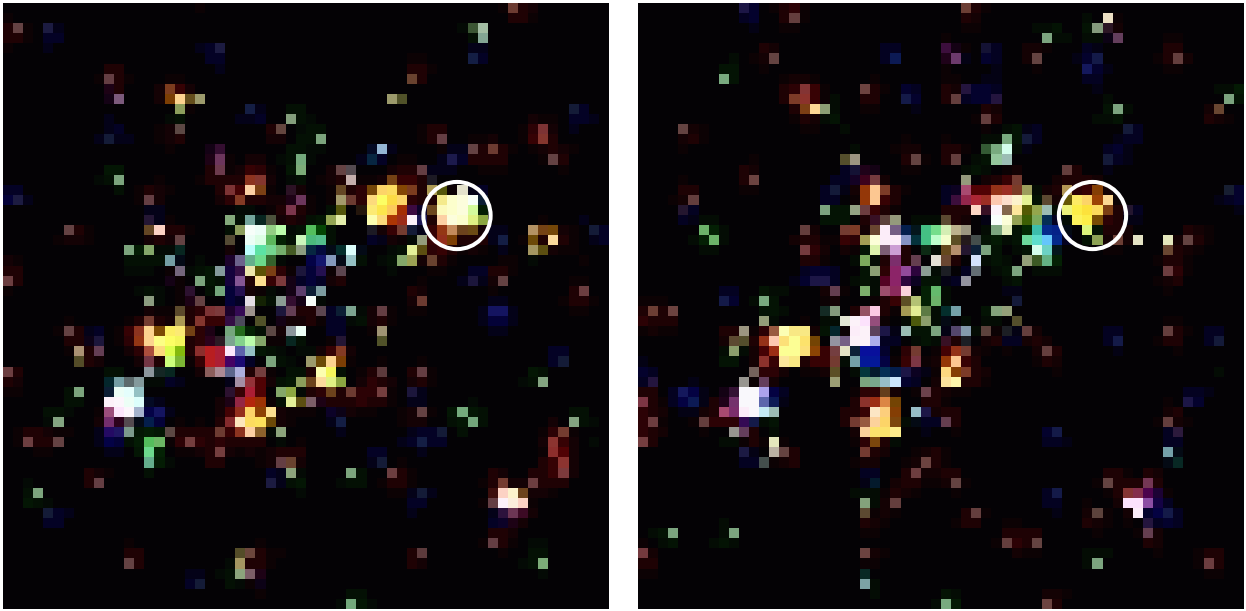}
\caption{Images of the globular cluster NGC~6440.  The left panel shows the data obtained in
  the 2000 July 4 \textit{Chandra} observation, where as the right panel shows the new
  \textit{Chandra} observation from 2003 June 26. The location of CX1 is marked with a circle.
  Both images show the neutron star X-ray
  transient CX1 in a quiescent state.  They are both plotted on the same scale
  ($29.5^{\prime\prime}\times29.5^{\prime\prime}$) for direct
  comparison.  East is left and north is upward.  The red color is for the 0.3-1.5 keV energy
  range, green for 1.5-2.5 keV, and blue for 2.5-8.0 keV.}
  \label{fig:NGC6440}
\end{figure}

A color image of the cluster was produced from the 2003 June 26 observation (hereafter
observation 2; see Fig.\ref{fig:NGC6440}, right panel) and a similar image was produced from
the 2000 July 4 observation for comparison (observation 1; see Fig. \ref{fig:NGC6440}, 
left panel).  It can be seen that during both of these observations, the source was in a
quiescent state \citep[see][for a comparison of observation 1 with the
August 2001 \textit{Chandra} observation of the source in outburst]{2001ApJ...563L..41I}.  In
this figure, the red color is for the 0.3-1.5 keV energy
range, green for 1.5-2.5 keV, and blue for 2.5-8.0 keV.  When
creating these images we removed the pixel randomization that is added in the standard data
processing, this slightly enhances the spatial resolution of the images.
We used `wavdetect' to detect the point sources in the cluster and determine their positions.
To determine that we
had correctly identified CX1 in the new observation we calculated the coordinate offset
between the two observations of the five brightest sources detected 
\citep[CX2 - CX6, as named in][]{2002ApJ...573..184P}.  The mean offset in RA =
$0.39^{\prime\prime}$ and in Dec =
$-0.07^{\prime\prime}$ with standard deviations of these offset being $0.11^{\prime\prime}$ in
RA and $0.13^{\prime\prime}$ in Dec.  The offset of CX1 between the two observations was
$0.41^{\prime\prime}$ in RA and $-0.04^{\prime\prime}$ in Dec.  These offsets are well within
one standard deviation of the averaged offsets measured for the other sources,
hence we conclude that we have detected the same source in both observations.

\subsection{Spectral Analysis}\label{sec:specan}

Figure \ref{fig:NGC6440} shows a change in color of the source between the two observations. 
The more yellowish X-ray colors of the second observation suggest that the spectrum is softer
in the second observation.  Since the launch of \textit{Chandra} there has been continuous
degradation of the ACIS quantum
efficiency\footnote{See http://cxc.harvard.edu/cal/Acis/Cal\_prods/qeDeg/} which is most
severe at lower energies.  The effect of this degradation would be to make the second
observation harder, and hence this actually strengthens our impression based on a comparison of
the two images that the spectrum of the second observation is softer.

We extracted the count rates using a circle of radius $1.5^{\prime\prime}$ centred on the
source position as the source extraction region and an annulus from $17^{\prime\prime}$ to
$28^{\prime\prime}$ centred on the cluster centre as the background region.  No sources were
detected in this annulus.  For the source CX1 we detected 247 photons for observation 1 (for the
 photon energy range 0.3 - 10 kev)\footnote{A total number of 251 photons are detected for
observation 1 in the full energy range of Chandra, as previously found in
\citet{2001ApJ...563L..41I}.} with 0.7 background photons
giving a net count rate of $0.0106 \pm 0.0007$ counts s$^{-1}$ compared to 108 photons for
observation 2 (also for the photon energy range 0.3 - 10 keV), 0.6 background photons, and a net
count rate of $0.0045 \pm 0.0004$ counts s$^{-1}$. 
Clearly, we see a difference in the net count rates between the two observations strongly
suggesting that the source was variable between the two quiescent observations.  To further
investigate the spectral and count rate variations we observed, we extracted the source spectra
using the randomized data and the same extraction regions as above by using the CIAO tool
`psextract'.  This also creates the response matrices and ancillary response files, and
automatically corrects the latter for the degradation of the low-energy quantum efficiency of the
CCD as mentioned above.
The spectrum was grouped into bins of 10 counts.  Greater than 15 - 20 counts per bin are
formally required to use the $\chi^2$ statistic, but we decided on this number because the 2nd
observation had few net counts.  We checked our results by using the Cash statistics
\citep{1979ApJ...228..939C} and found the results to be consistent with the $\chi^2$ method
(within the 90 per cent confidence levels);
we only present results obtained using $\chi^2$ statistics.
In our spectral analysis we use the neutron star hydrogen atmosphere (NSA) model for weakly
magnetised neutron stars \citep*{1996A&A...315..141Z}.  Various other models, including
blackbody and disk blackbody models, fit the data satisfactorily but the NSA model is the most
commonly used and currently accepted to be a better physical description of
the emission originating from neutron star surfaces than other models \citep[see e.g.][for an
in depth discussion]{1999ApJ...514..945R,2000ApJ...529..985R}.

\subsubsection{Individual Spectral Fits}\label{sec:sepfits}
 
Initially we fitted the two spectra separately using Xspec version 11.2.0
\citep{1996adass...5...17A}.  For observation 1 the model used consists of an absorbed
neutron star atmosphere
model plus power-law.  The inclusion of a power-law to account for the hardest photons
significantly improves the fit.  For observation 2, however, the inclusion of a power-law does
not improve the fit and, in fact, when a power-law is included, Xspec forces the value of the
power-law index to a high, unphysical value.  For example, when assuming a distance
to the source of $d = 8.1$, 8.5 or 8.9 kpc, the best fitting power-law index is 9.5 in each case.
An absorbed NSA model excluding a power-law is therefore used when fitting the spectrum of
observation 2.

The normalization in the NSA model \citep{1996A&A...315..141Z} is given as $1/d^2$ where
\textit{d} is the distance in parsecs.
When leaving the normalization of the NSA model free this parameter could not be constrained
well, with values between $9.2\times10^{-7}$ and
$1.2\times10^{-10}$ and a best fit value of $5.0\times10^{-9}$. This corresponds to a
distance range of 1 - 93 kpc with a best fit value of 14 kpc.  Leaving the NSA normalization free
leads to large uncertainties in the other parameters too.
The best known distance to NGC~6440 from optical observations is well defined as being
$8.5 \pm 0.4 \mathrm{kpc}$ \citep{1994A&AS..108..653O} which is consistent with our range of
distances.  By fixing the NSA normalization, the uncertainties in the other parameters were
reduced.  We set the NSA normalization using $d$ = 8.1, 8.5 and 8.9
kpc, fully covering the allowed distance range from the optical observations.  A `canonical'
neutron star with a mass of 1.4 $\mathrm{M}_\odot$ and a radius of 10 km was also assumed.
Therefore, the only allowed free parameter in the NSA model was the temperature.  When fitting
to the individual observations the column density and the power-law spectral index and
normalization were free parameters.  The results of the spectral fits to the
individual observations are shown in Tables \ref{tab:nsaobs1} and \ref{tab:nsaobs2}. From
these results it can be seen that the different NSA normalization values do not greatly
affect the fit parameters.  The flux of the source is seen to change between the two
observations.  The model was extrapolated to the energy range 0.01-100 keV to give an estimate
of the bolometric flux.  Between the two observations the 0.5-10 keV flux decreases by a factor
of $\sim$2.  To determine the
errors in the fluxes, each free fit parameter was fixed to its minimum or maximum value in turn,
and only one at a time with the exception of the power law component.  This component was fixed
to its best fitting value as it is highly unconstrained and if left free gives unreasonably large
errors (e.g. a factor of approximately $10^2$).  The model was then refitted to the data and new
flux values calculated.  Once this was done for every free parameter, the total flux range
was used to give the flux errors.

\subsubsection{Combined Spectral Fits}\label{sec:combfits}

\begin{figure}
\plotone{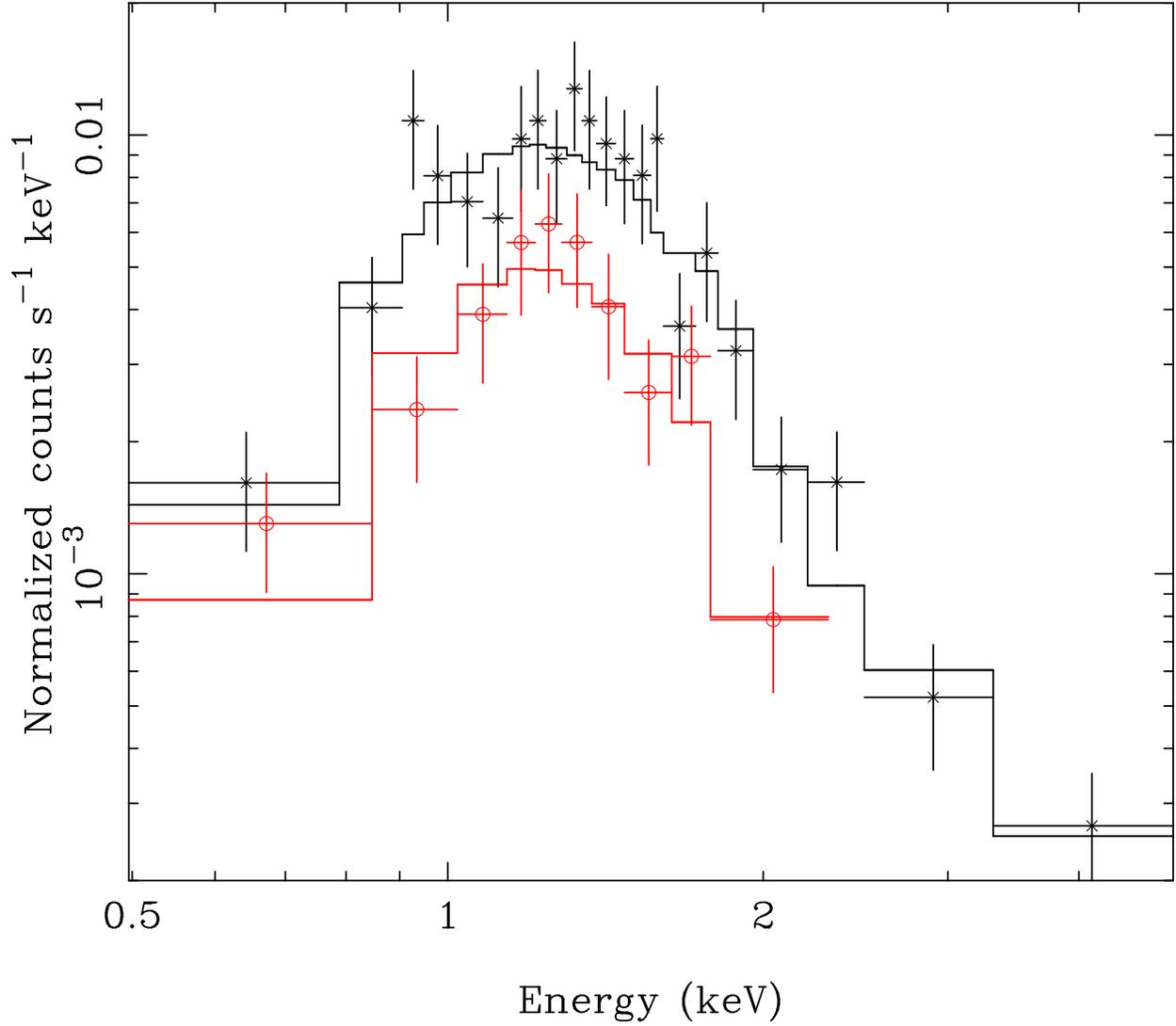}
\caption{X-ray spectra of the X-ray transient neutron star in NGC~6440 during
   observation 1 (crosses) and observation 2 (circles).  The solid lines through the
   data points are the best fitting models to the data.}
  \label{fig:ngc6440_spectra}
\end{figure}

To further investigate the cause behind the variation in flux we fit the two
observations simultaneously (Table \ref{tab:nsaboth}).
The spectra and fitted models can be seen in Fig. \ref{fig:ngc6440_spectra}.
As we expect the column density to be very similar for each observation, it was decided to tie
this parameter between observations.  Again a canonical neutron star was used and only the
temperature in the NSA model was left as a free parameter.  The power-law spectral index and
normalization for the first observation were allowed to be free, where as the normalization was
initially fixed to zero for the second observation.
After the model had been fit to the data, the power-law spectral index of the
second observation was fixed to the value obtained for the first observation and the 90 per
cent confidence limit on the normalization was determined. This
gives us an indication of the upper limit of the power-law normalization for the second
observation, assuming that the power-law spectral index was the same for both observations. 
In this case, taking the distance to the source to be 8.5 kpc, we get an upper limit on the
power-law normalization of $1.7\times10^{-5}$, with the power-law index being 2.5.  This gives a
corresponding upper limit on the
unabsorbed 0.5-10 keV flux of $1.0\times10^{-13}$ erg cm$^{-2}$ s$^{-1}$ and hence the maximum
contribution of the power law component to the 0.5-10 keV flux of 10 per cent.  Similar upper
limits were determined when taking $d = 8.1$ and 8.9 kpc (see Tab. \ref{tab:nsaboth}).
Power-law indices that are lower than our observed value
for the first observation have been measured in other systems \citep[e.g.][]{1999ApJ...514..945R},
and it is possible that the power-law index could have changed between the observations.
Therefore, to investigate whether the power-law index or normalization is variable between the two
observations the upper limits were also determined when setting the power-law spectral index of
the second observation to 0.5 and 1.0.
The corresponding upper limits in the power law
normalization were $2.9\times10^{-6}$ and $4.3\times10^{-6}$ respectively when taking the distance
of the source to be 8.5 kpc.  This gives unabsorbed 0.5-10 keV fluxes of $1.6\times10^{-13}$ and
$1.3\times10^{-13}$ erg cm$^{-2}$ s$^{-1}$ and the maximum contributions of the power-law
component to the 0.5-10 keV flux of 44 and 31 per cent, respectively.  As the upper-limit to the
power-law normalization for observation 2 is strongest when the power-law has a spectral index
equal to that of the first observation, it is possible that the power-law index remains unchanged
between the observations and it is the normalization that has changed, though we cannot be
conclusive. 

\begin{figure}
\plotone{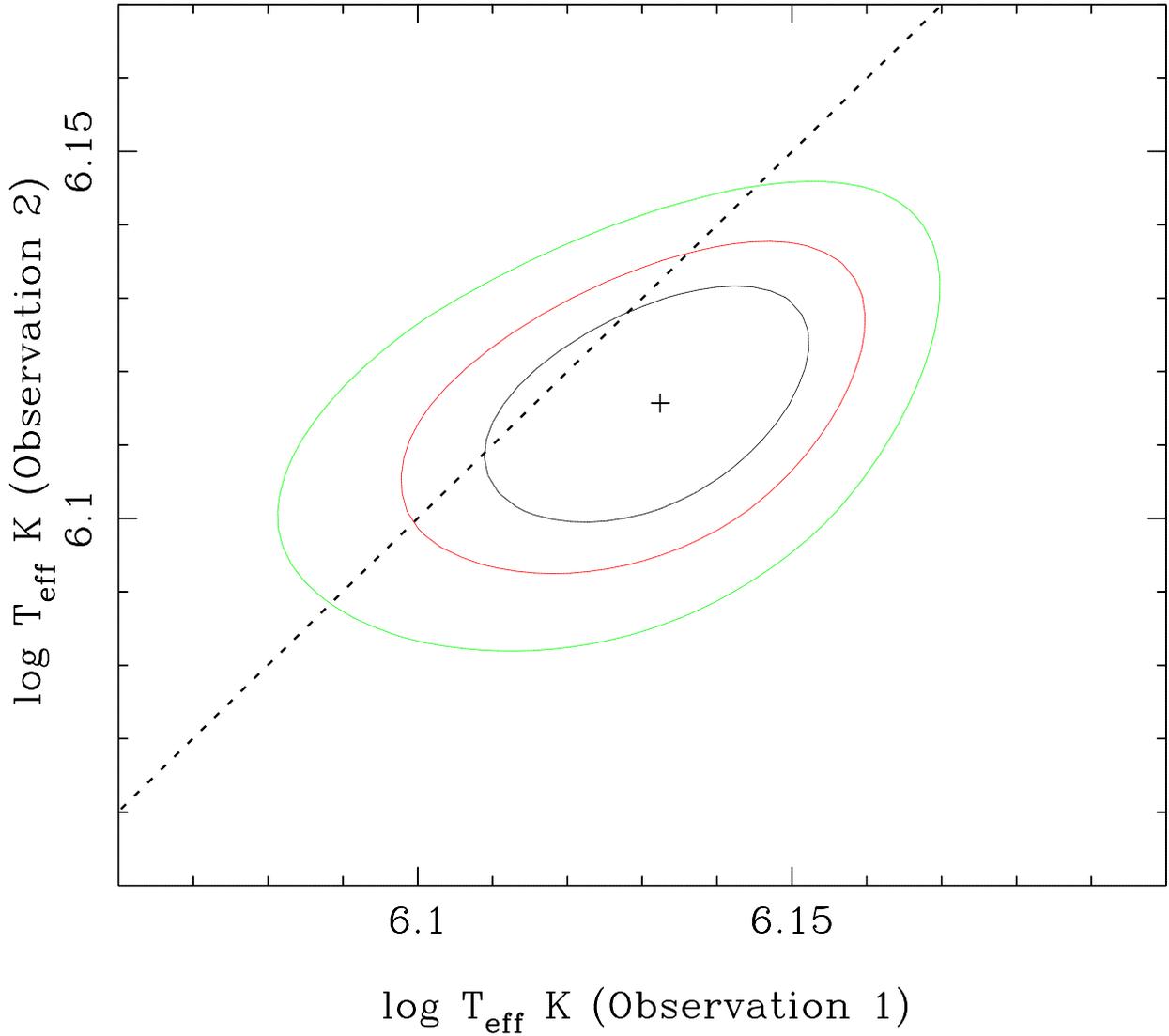}
\caption{Comparison of the effective temperature (for an observer on the surface of the neutron
  star) between observations 1 and 2 when taking the distance to the source to be 8.5 kpc.  The
  contours mark the 68.3\%, 90\%, and 99\% confidence levels.  The dashed line indicates where
  the temperature of observation 1 equals the temperature of observation 2.  The contours were
  determined with the power-law fixed to the best fitting value.}
  \label{fig:contours}
\end{figure}

Assuming that the distance to the source is 8.5 kpc, we found an
unabsorbed 0.5-10 keV flux of $1.9 \pm 0.3\times10^{-13}$
erg cm$^{-2}$ s$^{-1}$ in observation 1 and $0.9 \pm 0.2 \times10^{-13}$ erg cm$^{-2}$
s$^{-1}$ in observation 2.  Our results show that the
unabsorbed 0.5-10 keV flux reduces by a factor of $2.1 \pm 0.6$ between the first and
second observation.  This confirms our conclusion (section \ref{sec:specan}) that the neutron
star X-ray transient in NGC~6440 exhibits variability in quiescence.
The contribution to the bolometric flux from the NSA component was
found to be $1.6 \pm 0.4 \times10^{-13}$ erg cm$^{-2}$ s$^{-1}$ in observation 1 whereas all
the bolometric flux comes from the NSA component in observation 2 and is found to be
$1.4 \pm 0.2 \times10^{-13}$ erg cm$^{-2}$ s$^{-1}$.  So, the NSA component of the bolometric
flux is seen to be consistent between the two observations.  The corresponding effective
temperatures of the thermal component for the two observations
are seen to remain constant to within the 68.3\% confidence level.  This is illustrated in
Fig. \ref{fig:contours} where the line of equal temperature between the observations goes within
this 68.3\% confidence contour.

\section{DISCUSSION}

We have presented a new \textit{Chandra} observation of the globular cluster NGC~6440 during
a time when the neutron-star X-ray transient CX1 was in a quiescent state and compared this
observation with a previous \textit{Chandra} observation also taken whilst CX1 was in a
quiescent state.  Both spectra have been acceptably fitted with NSA models using a `canonical'
neutron star at the distance of NGC~6440 combined with galactic absorption.  In observation 1
the addition of a power-law component improves the fit at higher energies, as has previously
been found for other quiescent neutron stars (e.g. Aql X-1 \citep{2001ApJ...559.1054R,
2002ApJ...577..346R,1998ApJ...499L..65C}, Cen X-4 
\citep{2001ApJ...551..921R,2004ApJ...601..474C}), but the
addition of such a term to the second observation does not improve the fit.   We have shown
that the 0.5-10 keV flux is seen to decrease by a factor of $\sim$2 between these
two observations and that the neutron star atmosphere component to the bolometric flux is seen to
be consistent.  Our results for observation 1 are found to be consistent with the previous
analysis by \citet{2001ApJ...563L..41I}.

Other quiescent neutron stars which have been observed to be variable during quiescence include
Aql X-1, Cen X-4, MXB 1659-29 and KS 1731-260. 
\citet{2001ApJ...559.1054R,2002ApJ...577..346R} account the variability in Aql X-1 to a change
in the neutron star effective temperature caused by variable residual accretion onto the neutron
star, whereas \citet{2003ApJ...597..474C} prefer to explain the variability due to correlated
variations of the power-law component and the local column density, supporting the idea of shock
emission.  The variability in Cen X-4 has been attributed to a changing power-law component
\citep{2001ApJ...551..921R} or a variable local column density combined with variation in either
the soft or hard spectral component, or both \citep{2004ApJ...601..474C}.
Long accretion events in MXB 1659-29 and KS 1731-260 may have heated the crust of the neutron
stars in
these systems considerably out of thermal equilibrium with the cores and the quiescent
variability can be explained by the cooling of the crusts toward renewed equilibrium with the
cores \citep{2002ApJ...573L..45W,WijnandsMXB}.  However, the outbursts of the transient in
NGC~6440 are short (compared to the outbursts of MXB 1659-29 and KS 1731-260) during which the
neutron star crust is only slightly heated and soon
\citep[within weeks; e.g.][]{1998ApJ...504L..95B} after the end of the outbursts, the crust
will be in thermal equilibrium with the core.  Since our quiescent observations were both taken
$\sim$2 years after the end of the outbursts, we expect that the crust has already reached
equilibrium with the core and therefore a cooling crust can probably not explain the observed
variability.

From our spectral analysis of CX1, the significant difference between the observations to note is
the undetectability of a power-law component in the second observation.
Although a hard power-law component is seen in observation 1, it is not detected in
observation 2.  Both observations were of the same exposure and so will be equally sensitive to
the hard tail.  For the photon energy range 3-10 keV, 15.7 net counts were detected from the
source in the first observation (0.3 background counts), whereas no counts were detected in this
energy range in second observation.  Clearly, there is a definite change in the power-law
component between these observations.  But, it is unclear from these results whether this is due
to a variable spectral index or normalization, or both.  This change in property, or properties,
of the power-law component likely accounts for almost all of the change in luminosity.
The unchanging neutron star effective temperature between the
observations supports the idea that the thermal quiescent luminosity is set by deep crustal
heating \citep{1998ApJ...504L..95B}.
 
We have clearly shown that the neutron star X-ray transient in NGC~6440 was variable between two
quiescent observations.  This variability is attributable to a change in the power-law
component whilst the thermal component is consistent with remaining constant.  This is similar to
the results found
for Aql X-1 \citep{2003ApJ...597..474C} and Cen X-4 \citep{2004ApJ...601..474C} where the thermal
component from the neutron star is consistent with remaining constant and the power-law is seen
to be variable.  These
observations all indicate that there is a stable thermal flux coming from the neutron star
surface, likely set by deep crustal heating, and a variable flux at higher energies.
Although the nature of the power-law component is uncertain, one proposal is that
it could be due to matter interacting with the magnetic field of the neutron star in some way
\citep[e.g.,][]{1998A&ARv...8..279C,2000ApJ...541..849C}.
In particular,   
\citet{2003ApJ...597..474C} and \citet{2004ApJ...601..474C} explain variable power-law flux they
observed as emission at the shock between a pulsar wind and a variable amount of inflowing matter
from the companion star.  They base this on the fact that they found evidence that the spectral
variability in Aql X-1 and Cen X-4 could be interpreted to correlated
variations of the power-law component and the local column density. Unfortunately, we
were unable to determine if a similar correlation in CX1 was underlying the spectral
variability we found in this source, as we cannot fit a power-law to both observations.  However,
some variation in accretion rate onto the magnetosphere or variation in the interaction of the
pulsar wind with the matter still outflowing from the companion star could, in principle, explain
the variability observed in CX1, with less matter interacting with the neutron star magnetic
field during observation 2 compared to observation 1.
If there is residual accretion occuring it is unlikely to accrete down to the surface
since that would produce thermal-like emission \citep{zampieri1995}, due to the heating of the
surface, over the thermal flux expected from a cooling neutron star.

\begin{figure}
\plotone{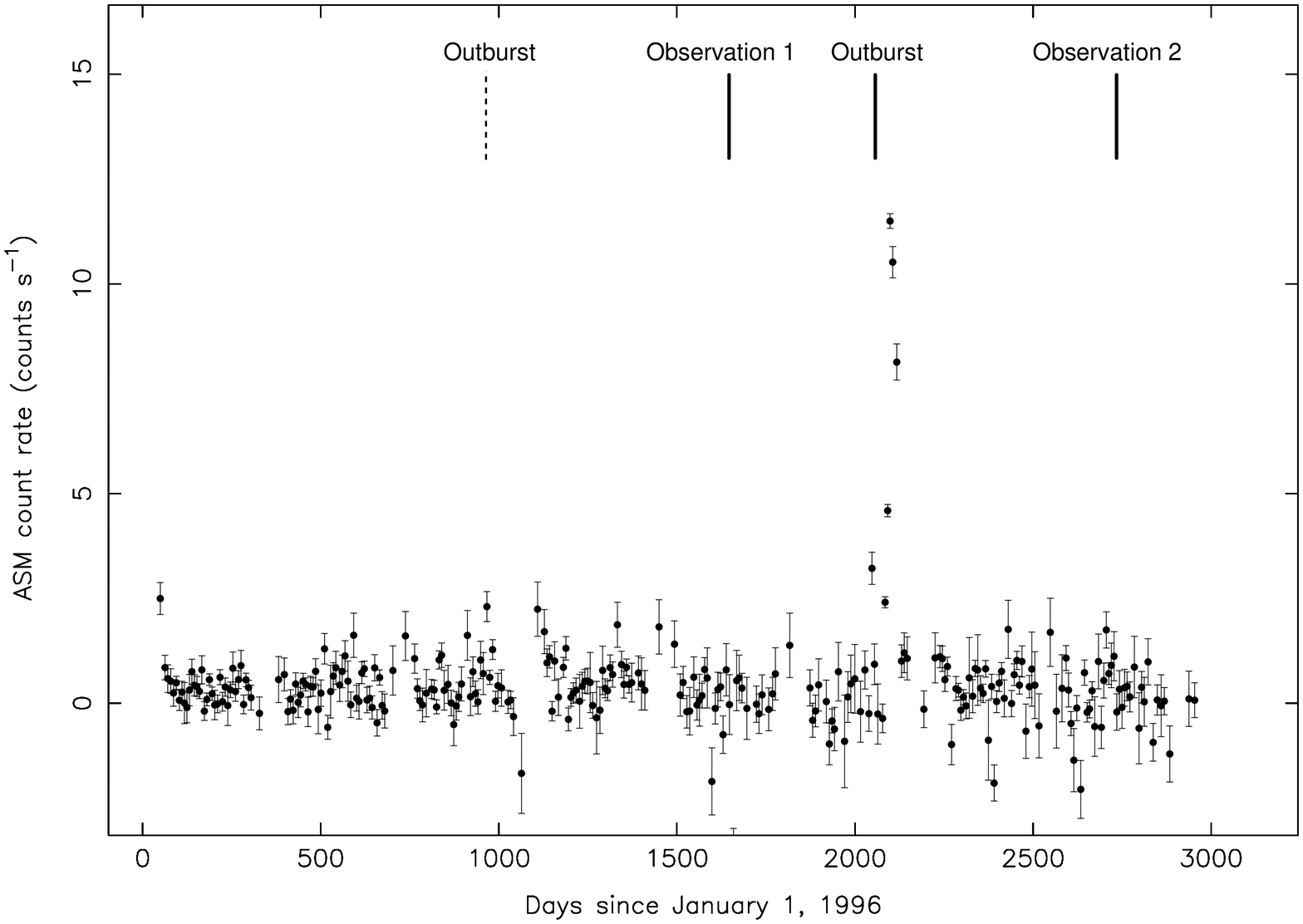}
\caption{\textit{RXTE} ASM light curve of the transient in NGC 6440.  Each point is
  averaged over seven days.  The dates of the three \textit{Chandra} observations are marked
  with solid lines.  It can be seen that both observation 1 and observation 2 were taken whilst
  the neutron star X-ray transient in NGC~6440 was in a quiescent state.  The dashed line
  indicates the 1998 August outburst.}
  \label{fig:asm}
\end{figure}

The transient in NGC 6440 was seen to be in a quiescent state since the outburst
in 1971 until the 1998 outburst, and then until the 2001 outburst.  We can use this information to
predict the thermal flux during quiescence.  From the
time averaged accretion flux, $\langle F_{acc}\rangle$, the expected thermal flux in quiescence,
$F_q$, can be
predicted using the \citet{1998ApJ...504L..95B} model and assuming standard core
cooling: $F_q = \langle F_{acc}\rangle/135$
\citep{1998ApJ...504L..95B,wijnands2001,2002ApJ...577..346R}.
From this, it can be derived \citep{wijnands2001} that 
$F_q \sim t_o/(t_o + t_q) \times \langle F_o \rangle/135$, where $\langle F_o\rangle$
is the time-averaged flux
during outburst, $t_o$ is the time the source was in outburst, and $t_q$ the averaged
time the source is in quiescence.  The time-averaged flux during outburst for CX1
can be estimated from the \textit{RXTE}/ASM lightcurve (see Fig. \ref{fig:asm}).
To convert the ASM count rate to
a flux, we use WebPIMMS, modelling the outburst spectrum as a power-law with
index 1.7 and $N_H = 1.2\times 10^{22}$
\citep[valid for the energy range 1-40 keV;][]{1999A&A...345..100I}.  This gives the
time-averaged flux during outburst as $\langle F_o\rangle = 5.79\times10^{-9}$
erg cm$^{-2}$ s$^{-1}$
in the energy range 1-40 keV.  However, the bolometric flux could be a factor of a few higher
than this depending on the correct spectral model and energy range. Assuming that
no other outbursts were detected since the 1971 outburst, we get the total time
in quiescence up until the 2001 outburst as $t_q$ = 10973 days, estimating the total
outburst time, $t_o \sim 43$ days (from the ASM lightcurve).  The predicted quiescent
flux is therefore, $F_q = 1.7\times10^{-13}$ erg cm$^{-2}$ s$^{-1}$.  Although the
bolometric correction
could make this predicted flux a factor of a few higher, the uncertainties in our
assumptions are likely to be large, making this predicted flux in good agreement
with the observed  
quiescent flux of the NSA component that we found from the Chandra observations (Tab. 1-3).
This shows that standard cooling model can account for the observed behaviour of
CX1.  Other quiescent neutron stars cannot easily be explained by the standard
cooling model and require enhanced core cooling, for example, KS1731-260
\citep{wijnands2001} and Cen X-4 \citep{2001ApJ...551..921R}, in which case the neutron stars
in these systems could be more massive ($> 1.7$ M$_{\odot}$) than those in prototypical neutron
star transients \citep[e.g.][]{colpi2001}. 
Further monitoring of this, and other, neutron star transients in a quiescent state will better
constrain the quiescent properties and help determine the cause of the observed variability.  

\acknowledgements

EMC acknowledges the support of a PPARC Studentship at the University of St Andrews.

\clearpage

\begin{deluxetable}{lccc}
\tabletypesize{\small}
\tablecolumns{4} 
\tablewidth{0pc}
\tablecaption{Spectral results for observation 1}
\tablehead{
   & \multicolumn{3}{c}{Assumed Distance to NGC 6440} \\
  Model Parameter & 8.1 kpc & 8.5 kpc & 8.9 kpc}
\startdata
$N_H \left(10^{22}\, \mathrm{cm}^{-2}\right)$ & $0.7 \pm 0.1$ 
& $0.7 \pm 0.1 $
& $0.7 \pm 0.1 $\\

$kT_{\mathrm{eff}}^{\infty}$ (eV) & $88 \pm 5$
& $91 \pm 5$
& $92 \pm 5$\\

Power-law index & $2.5 \pm 1.0$
& $2.4 \pm 1.0$
& $2.4 \pm 1.0$\\

Power-law normalization ($10^{-5}$) & $2.2^{+3.5}_{-1.4}$ 
& $2.0^{+3.4}_{-1.3}$ 
& $1.9^{+3.3}_{-1.3}$\\

Flux (0.5 - 10 keV) ($10^{-13}$ ergs cm$^{-2}$ s$^{-1}$)
& $1.9 \pm 0.4$
& $1.8 \pm 0.4$
& $1.8 \pm 0.4$\\

Fraction of 0.5 - 10 keV flux in power-law
& $0.41 \pm 0.2$
& $0.40 \pm 0.2$
& $0.40 \pm 0.2$ \\

NSA contribution to the bolometric flux ($10^{-13}$ ergs cm$^{-2}$ s$^{-1}$)
& $1.7 \pm 0.4$
& $1.7 \pm 0.4$
& $1.6 \pm 0.4$\\

$\chi_\nu^2$ & 0.74 & 0.73 & 0.73\\

\enddata
\label{tab:nsaobs1}
\tablecomments{The error bars represent 90$\%$ confidence levels.  For the NSA model, we used a
neutron star mass of 1.4 M$_\odot$, a radius of 10 km and the neutron star hydrogen atmosphere
model for weakly magnetized neutron stars of \citet{1996A&A...315..141Z}.}
\end{deluxetable}

\begin{deluxetable}{lccc}
\tabletypesize{\small}
\tablecolumns{4} 
\tablewidth{0pc}
\tablecaption{Spectral results for observation 2}
\tablehead{Model Parameter
   & \multicolumn{3}{c}{Assumed Distance to NGC 6440} \\
    & 8.1 kpc & 8.5 kpc & 8.9 kpc}
\startdata
$N_H \left(10^{22}\, \mathrm{cm}^{-2}\right)$ & $0.7 \pm 0.2$ 
& $0.7 \pm 0.2$
& $0.7 \pm 0.2 $\\

$kT_{\mathrm{eff}}^{\infty}$ (eV) & $85 \pm 5$
& $87 \pm 5$
& $88 \pm 5$\\

Flux (0.5 - 10 keV) ($10^{-13}$ ergs cm$^{-2}$ s$^{-1}$)
& $0.9 \pm 0.3$
& $0.9 \pm 0.2$
& $0.9 \pm 0.2$\\

Bolometric flux ($10^{-13}$ ergs cm$^{-2}$ s$^{-1}$)
& $1.5 \pm 0.3$
& $1.4 \pm 0.3 $
& $1.4 \pm 0.3$\\

$\chi_\nu^2$ & 0.66  & 0.66 & 0.65 \\
\enddata
\label{tab:nsaobs2}
\tablecomments{The error bars represent 90$\%$ confidence levels.  For the NSA model, we used a
neutron star mass of 1.4 M$_\odot$, a radius of 10 km and the neutron star hydrogen atmosphere
model for weakly magnetized neutron stars of \citet{1996A&A...315..141Z}.}
\end{deluxetable}

\begin{deluxetable}{lccc}
\tabletypesize{\small}
\tablecolumns{4} 
\tablewidth{0pc}
\tablecaption{Spectral results when simultaneously fitting to observations 1 and 2}
\tablehead{

  & \multicolumn{3}{c}{Assumed Distance to NGC 6440} \\            
  Model Parameter & 8.1 kpc & 8.5 kpc & 8.9 kpc\\ \tableline
\multicolumn{4}{c}{Parameters for obs. 1}}

\startdata
$N_H \left(10^{22}\, \mathrm{cm}^{-2}\right)$ & $0.7 \pm 0.1$ 
& $0.7 \pm 0.1$
& $0.7 \pm 0.1$\\

$kT_{\mathrm{eff}}^{\infty}$ (eV) & $87 \pm 5$
& $89 \pm 4$
& $92 \pm 5$\\

Power-law index & $2.6 \pm 0.8$
& $2.5 \pm 0.8$
& $2.4 \pm 0.9$\\

Power-law normalization $(10^{-5})$ & $2.7^{+2.7}_{-1.5}$ 
& $2.3^{+2.5}_{-1.3}$ 
& $2.0^{+2.3}_{-1.2}$\\

Flux (0.5 - 10 keV) ($10^{-13}$ ergs cm$^{-2}$ s$^{-1}$)
& $1.9 \pm 0.3$
& $1.9 \pm 0.3$
& $1.8 \pm 0.3$\\

Fraction of 0.5 - 10 keV flux in power-law
& $0.46 \pm 0.2$
& $0.43 \pm 0.2$
& $0.40 \pm 0.2$ \\

NSA contribution to the bolometric flux ($10^{-13}$ ergs cm$^{-2}$ s$^{-1}$)
& $1.6 \pm 0.4$
& $1.6 \pm 0.4$
& $1.6 \pm 0.3$\\

\tableline
\multicolumn{4}{c}{Parameters for obs. 2} \\ \tableline

$kT_{\mathrm{eff}}^{\infty}$ (eV) & $85 \pm 4$
& $86 \pm 4$
& $88 \pm 4$\\

Power-law normalization $(10^{-5})$ & $<2.0$ 
& $<1.7$ 
& $<1.5$\\

Flux (0.5 - 10 keV) ($10^{-13}$ ergs cm$^{-2}$ s$^{-1}$)
& $0.9 \pm 0.2$
& $0.9 \pm 0.2$
& $0.9 \pm 0.2$\\

Bolometric flux ($10^{-13}$ ergs cm$^{-2}$ s$^{-1}$)
& $1.5 \pm 0.2$
& $1.4 \pm 0.2$
& $1.4 \pm 0.2$\\ 
\tableline

$\chi_\nu^2$ & 0.69  & 0.68  & 0.68 \\
\enddata
\label{tab:nsaboth}
\tablecomments{The error bars represent 90$\%$ confidence levels.  For the NSA model, we used a
neutron star mass of 1.4 M$_\odot$, a radius of 10 km and the neutron star hydrogen atmosphere
model for weakly magnetized neutron stars of \citet{1996A&A...315..141Z}.}
\end{deluxetable}

\end{document}